\documentclass[%
reprint,
amsmath,amssymb,
aip,
jcp
]{revtex4-1}

\usepackage{graphicx}
\usepackage{dcolumn}
\usepackage{bm}
\usepackage{hyperref}
\usepackage[english]{babel}
\usepackage{color}
\usepackage{epstopdf}

\begin{document}

\title{\color{blue}Complex crystalline structures in a two-dimensional core-softened system}

\author{N.P. Kryuchkov}
\author{S.O. Yurchenko}
\email{st.yurchenko@mail.ru}
\affiliation
{Bauman Moscow State Technical University, \\
105005 2nd Baumanskaya str. 5, Moscow, Russia}
\author{Yu. D. Fomin}
\email{fomin314@gmail.com}
\author{E. N. Tsiok}
\author{V. N. Ryzhov}
\email{ryzhov@hppi.troitsk.ru}
\affiliation
{Institute for High Pressure Physics RAS, 108840 Kaluzhskoe shosse, 14, Troitsk, Moscow,
Russia}



\date{\today}

\begin{abstract}
A cascade of phase transitions from square to hexagonal lattice is studied in 2D system of particles interacting via core-softened potential.
Due to the presence of two length-scales of repulsion, different local configurations with four, five, and six neighbors enable, leading to formation of complex crystals.
The previously proposed interpolation method is generalized for calculation of pair correlations in crystals which elemental cell consists of more than one particle.
A high efficiency of the method is illustrated using a snub square lattice as a representative example.
Using molecular dynamics simulations, it is found that the snub square lattice is being broken under heating,
generating high density quasicrystalline phase with 12-fold symmetry.
Simple theoretical model is proposed to explain the physical mechanism governing this phenomenon:
With density growth (from square to hexagonal phases), the concentrations of different local configurations randomly realized through plane tilling are being changed that minimizes the energy of the system.
The calculated phase diagram in the intermediate region of densities justifies the existence of HD12 phase and demonstrates a cascade of the first-order transitions ``square -- HD12 -- hexagonal'' solid phases with the density growth.
The results allow us to better understand the physical mechanisms responsive for formation of quasicrystals, and, therefore, should be of interest for broad community on material science and soft matter.
\end{abstract}


\maketitle


\section{Introduction}

\section{Introduction}

The behavior of two-dimensional (2D) condensed systems attracts interest for interdisciplinary studies in physical chemistry, materials science, and soft matter \cite{book:1606888, ivlev.book}.
Two-dimensional and quasi-2D (confined or absorbed) systems of interacting particles are widely present in systems with developed surfaces and/or interfaces,
since such gradient areas are appealing for ions and colloidal particles \cite{PhysRevLett.114.108301, Poulichet12052015, PhysRevE.95.022602, doi:10.1021/acs.langmuir.6b01644}.
Moreover, 2D and quasi-2D systems are of technological importance because their use in the synthesis of advanced photonic materials is promising\cite{C6SM00031B, doi:10.1063/1.3115641, Vardeny2013}.
In particular, 2D colloidal crystals can be used as structured substrates and seeds for bulk photonic crystals employed for sensors and light conversions, in optical composites, and for spectroscopy using optical field localization (see Refs.\cite{doi:10.1063/1.4892363, doi:10.1063/1.4880299, 0022-3727-50-5-055105, 1468-6996-15-3-034805, doi:10.1021/acsnano.6b02400}).

Apart from applied significance, 2D systems have a long and instructive history of fundamental studies (see Refs.~\cite{kost2016,UFN2017,RevModPhys.89.040501} and references therein).
As early as in the 1930s, Landau and Peierls \cite{l1,p1,p2} showed that 2D crystals do not demonstrate a long-range positional order and corresponding mean-square displacements diverge logarithmically.
In spite of this, owing to a rather weak algebraical decay of positional correlations, typical experimental systems can be safely considered as crystals.
After the appearance of the Berezinskii--Kosterlitz--Thouless theory of phase transitions in the degenerate two-dimensional systems \cite{b1,b2,kt1}, a lot of attention was paid to the melting transition in two-dimensional systems.
In the widely accepted Berezinskii--Kosterlitz--Thouless--Halperin--Nelson--Young (BKTHNY) theory \cite{hn1,y1}, it is supposed that in contrast to the 3D case, where melting is always a first-order transition, 2D melting can occur through two continuous transitions with a new intermediate phase with a quasi-long-range orientational order, which is called the hexatic phase.
It can be stated with a high degree of confidence, which is based on real and computer simulations (see, e.g., \cite{UFN2017} and references therein), that systems with a long-range interaction (e.g., Coulomb and dipole--dipole interaction, and soft disks $1/r^n$ with $n\leq6$) are melted in accordance with the BKTHNY theory.
On the other hand, first-order melting without a hexatic phase in two dimensions is also possible \cite{c1,r1,r2,r3}.
Recently, another melting scenario was proposed \cite{kk1,kk2}.
According to this scenario, the crystal--hexatic phase transition is a continuous BKT-type transition and the hexatic--isotropic liquid phase transition is a first-order transition.
The last scenario seems true for some systems with short-range potentials \cite{kk1,kk2,hs,r4} (e.g., for soft disks, this scenario was shown to take place for $n>6$ \cite{kk2}).
At the same time, the microscopic mechanism of the first order hexatic-to-isotropic liquid transition is not clear.
The possible explanation can be based on the consideration of the dependence of the topological defect pairs unbinding on the core energy of this defect \cite{RN89}, but the further investigation of this issue is necessary.

At present, there are no precise theoretical criteria how to determine the crystal structure and melting scenario based on the form of the potential.
Along with the BKTHNY melting scenario, several possible melting mechanisms were suggested.
Phase transitions between fluids and solids can occur in 2D and quasi-2D systems, for instance, during heterogeneous crystallization \cite{doi:10.1021/jp073713d, doi:10.1063/1.4705393}, condensation on the surface and subsequent crystallization in the plane \cite{0953-8984-21-46-464114}, formation of clusters of particles on the fluid surface \cite{doi:10.1021/jp9085238, doi:10.1021/acs.jpcb.5b10105},
and in systems of attracting like-charged particles\cite{0034-4885-65-11-201, 0953-8984-12-8A-309}.

Until now, the most efforts have been focused on the study of two-dimensional (equilibrium and non-equilibrium) melting in triangular 2D crystals since they were the only lattice observed in experiments \cite{book:1606888, ivlev.book, PhysRevLett.103.015001, C0SM00813C, doi:10.1038/ncomms7942, doi:10.1038/natrevmats.2015.11, PhysRevE.96.043201}.
However, a lot of recent experimental studies demonstrate that other crystalline phases can appear in 2D materials with anisotropic interaction, e.g., graphene and square crystals of water \cite{sw} and iron \cite{sf,sf1}. Along with natural molecular systems, square and more complex lattices can be observed in 2D colloidal systems (confined by walls or gravitational forces).
For instance, numerous complex structures in the 2D Hertz system were reported \cite{C1SM05731F}, including snub square and Kagome lattices.
The square lattice can be formed due to complex interactions with attraction \cite{PhysRevE.74.021404, doi:10.1021/acs.jpcc.6b06704}.

Especially numerous structures have been observed recently in 2D systems with double-scale isotropic repulsive interactions, including the recent studies of quasicrystals \cite{POLB:POLB20537, IJCH:IJCH201100146, PhysRevLett.98.195502, Mikhael20042010, Mikhael2008, Dotera2014, C6SM01454B, doi:10.1038/nmat4963}.
Systems with this interaction type both in the form of a step-like potential \cite{doi:10.1063/1.4977934, C7SM00254H, CHUMAKOV2015279} and a softer version were studied.
Three-dimensional systems of particles with these interactions were also actively investigated \cite{doi:10.1063/1.2965880, PhysRevE.79.051202,  doi:10.1063/1.3668313, doi:10.1063/1.3643115, RCR2013} demonstrating rich variety of structures and properties, including polymorphism, maxima and minima on melting lines, glass transition, and water-like anomalies.
As was shown in Refs. \cite{doi:10.1063/1.3530790, doi:10.1063/1.4960113}, the diversity of stable structures can be yet increased by some generalization of this interaction type.

The double-scale repulsive interaction is of particular interest since it can serve as a simplified model of interaction between colloidal particles with polymeric shells~\cite{POLB:POLB20537, Fischer01022011, B601916C, LIKOS2001267}.
The promising methods of self-assembly could include the employment of tunable attraction produced by external rotating electric fields \cite{doi:10.1063/1.3115641, doi:10.1063/1.3241081, ADFM:ADFM201200400, s41598-017-14001-y, doi:10.1021/acs.jpcc.7b09317}.
These results can pave the way for controlled methods for synthesis of 2D complex colloidal structures\cite{ADFM:ADFM201002091}.
Another important issue is to visualize the motion of  colloidal microparticles using video microscopy.
The further particle-resolved analysis makes it possible to reveal generic physical mechanisms responsible for the formation of various structures in such systems and to understand the behavior of the many-body systems with different nature \cite{ivlev.book, book:1606888} since colloidal suspensions are known to be used as model systems for particle-resolved studies of molecular systems \cite{ivlev.book, C2SM26245B, C2SM26473K, doi:10.1038/srep28578, doi:10.1038/srep06132, doi:10.1038/ncomms7942, doi:10.1038/natrevmats.2015.11, doi:10.1038/ncomms14978}.

By the way, systems with designable interparticle interactions and tunable phase transitions provide a way for producing self-assembled materials for applications in photonics (sensors, optoelectronics, and light transformation), physical chemistry (catalysis, separation, membranes), and for fundamental particle-resolved studies of relations between the interaction kind, kinetics of self-assembly, assembled structures, and their properties.

In this work, we study a transition from a square to a high-density hexagonal lattice in a 2D system of particles interacting via a core-softened potential. Due to the presence of the two length scales corresponding to the core and shoulder repulsion,
the local configurations with four, five, and six neighboring particles are possible and the plane can be tiled by squares and triangles.
One of such five-neighbor tiling forms a snub square lattice whose unit cell consists of eight particles.
We study the snub square lattice using molecular dynamics (MD) simulations and the interpolation method (IM)\cite{doi:10.1063/1.4869863, doi:10.1063/1.4926945, 0953-8984-28-23-235401}, which are generalized here to calculate pair correlation functions in complex crystals with more than one particle per unit cell.
We find that pair correlations in the snub square lattice can be calculated in details analytically by the IM, but the lattice is metastable at low temperatures and is broken upon heating.
To understand the physical mechanism of this transition, we study the local configurations in intermediate structures.
We prove that the double degeneracy of the local five-bond configurations plays a crucial role in the intermediate density range,
providing a fluctuation mechanism for the formation of high-density quasicrystals with 12-fold symmetry (HD12 phase), instead of an expected snub square lattice.
Molecular dynamics  simulations and developed analytical approach (for zero-temperature limit) are employed to justify it and agree with each other.
The calculation of the phase diagram proves the existence of the quasicrystalline HD12 phase, which is located between the square and triangular phases, and remains to be stable until the melting line.
The results shed light to the physical mechanisms responsive for the cascades of phase transitions in 2D solids leading to the formation of quasicrystals, and, therefore, should be of interest for novel technologies of materials for applications in photonics and soft matter.

The Article is organized as follows:
In Sections \ref{RSQ-Methods-MD} and \ref{RSQ-Methods-Analysis}, we present details of our MD simulations.
In Section \ref{RSQ-Methods-IM}, we generalize for complex lattices the recently proposed interpolation method (IM), which enables the calculation of pair correlation functions in classical crystals.
In Section \ref{RSQ-Results-SnSq}, we analyze the snub square lattice, which appears to be dynamically stable as compared to square and hexagonal lattices.
However, the snub square lattice is broken irreversibly upon heating, which leads to the generation of a high-density quasicrystalline phase with 12-fold symmetry.
In Sec.~\ref{RSQ-Results-HD12}, we discuss the physical mechanism responsible for this phenomenon and present the corresponding calculated phase diagram.

\section{System and Methods}

\subsection{Molecular dynamics simulations}
\label{RSQ-Methods-MD}
We applied molecular dynamics (MD) simulations to study a 2D system of repulsive soft particles (RSPs) interacting via the following model potential:
\begin{equation}
  \varphi(r)/ \varepsilon = \left ( \frac{\sigma}{r} \right )^{14}+\frac{1}{2}\left(1-\tanh(k(r- \sigma_1))\right),
  \label{RSQ-eq1}
\end{equation}
where $\varepsilon$ is the interaction strength; $\sigma$ and $\sigma_1$ are the characteristic ranges of interaction with the particle core and its soft shoulder, respectively; and the index $k$ characterizes the smoothness of the repulsive shoulder at the particle periphery.
In this work, we used the parameters $\sigma_1 / \sigma =1.35$ and $k\sigma=10.0$,while the parameters $\varepsilon$ and $\sigma$ were employed to reduce the units for all the quantities, e.g.,
the areal density of particles is expressed as $\rho = N \sigma^2 /S$, where $N$ and $S$ are the total number of particles and the area of the system, respectively.

The MD simulations were performed for different temperatures in the same manner as was reported in Refs.\cite{1742-6596-510-1-012016, doi:10.1063/1.4896825, C4SM00124A, dftr2015}.
The system was simulated in the canonical ($NVT$) ensemble with $N=20808$ particles in a square box with periodic boundary conditions for $10^8$ steps with a time step of $\Delta=10^{-3}$ (in dimensionless units).
The first $3 \times 10^7$ steps were used for equilibration, while the last $7 \times 10^7$ steps were used to study the properties of the system.

\subsection{Analysis of MD results}
\label{RSQ-Methods-Analysis}
To analyze the system properties, we calculated the equation of state for different temperatures, the radial distribution functions,
the ground state energies in the zero-temperature limit and phonon spectra obtained using the Born--von Karman method\cite{Dove-book} for triangular, square, and snub square\cite{CHAVEY1989147} lattices.
For the detailed structural analysis of the system, we calculated the parameters of Voronoi cells in the system, in particular,
the coordination number for the local environment of each particle and the bond-orientational order parameter $\psi_m$:
\begin{equation}
   \psi_{i;m}=\left<\left|\frac{1}{\mathrm{NN}_i} \sum_{j=1}^{\mathrm{NN}_i} \exp(i m \theta_{ij})\right|\right>
   \label{RSQ-eq2}
\end{equation}
where  $\mathrm{NN}_i$ is the number of nearest neighbors of the $i$-th particle, $\mathbf{r}_{ij}=\mathbf{r}_j-\mathbf{r}_i$ is the radius vector between the $i$-th and $j$-th particles, $\theta_{ij}$ is the angle between $\mathbf{r}_{ij}$ and the $x$ axis, and $\left<...\right>$ denotes the averaging over all the particles in the system.
For all particles in an ideal triangular crystal, $\psi_6=1$ and $\psi_4=0$, whereas in an ideal square lattice, $\psi_6=0$ and $\psi_4=1$.

In addition to the $\psi_m$-order parameter, we used another order parameter used in Ref.\cite{Dotera2014} for the system as a whole:
\begin{equation}
 \chi_m=\left\langle\left|\frac{1}{N_{Bi}}\sum_{j}\exp\left(i m \theta_{ij}\right) \right|^2\right\rangle,
   \label{RSQ-eq3}
\end{equation}
where the summation is performed over all $j$-th particles located at the distance less than $\sigma_1$ from the $i$-th particle ($r_{ij}<\sigma_1$), and $N_{Bi}$ is the number of these particles.

To study the symmetry of the resulting structures, we calculated the structure factor by applying the Fourier transform as follows:
\begin{equation}
 S(\mathbf{q})=\frac{1}{N}\Big\langle\sum_j \exp\left(i \mathbf{q}\mathbf{r}_j\right)\Big\rangle_t,
\label{RSQ-eq4}
\end{equation}
where the summation is carried out over all particles $j$ in the system and $\langle\ldots\rangle_t$ denotes the time averaging.

\subsection{Interpolation method (IM)}
\label{RSQ-Methods-IM}

The interpolation method (IM) was elaborated in Refs.\cite{doi:10.1063/1.4869863, doi:10.1063/1.4926945, 0953-8984-28-23-235401} and makes it possible to calculate the pair correlation function $g(\mathbf{r})$ in classical crystals at finite temperatures.
Previously, the IM method was applied to simple crystals with one particle per unit cell.
In this work, we consider the snub square lattice as a representative example to illustrate the efficiency of the generalized IM.
Due to the complexity of the snub square lattice (eight particles per cell), we should use more general expressions for mean-square displacements (MSDs) than those used before\cite{doi:10.1063/1.4869863, doi:10.1063/1.4979325, doi:10.1063/1.4926945, 0953-8984-28-23-235401}.

In a crystal, the $g(\mathbf{r})$ function can be represented as the sum of independent correlation peaks $p_\alpha(\mathbf{r})$ corresponding to the contributions of different lattice sites $\alpha$\cite{doi:10.1063/1.4926945, 0953-8984-28-23-235401}:
\begin{equation}
g(\mathbf{r}) = \frac{V}{N}\sum_\alpha{p_\alpha(\mathbf{r}-\mathbf{r_\alpha})},
\label{RSQ-eq5}
\end{equation}
where $V$ and $N$ are the volume (area $S$ in the 2D case) and the total number of particles in the crystal, respectively;
$\mathbf{r}_\alpha$ is the equilibrium position of the site $\alpha$; and the summation is performed over all lattice sites.
The pair correlation peak $p_\alpha(\mathbf{r})$ for each site has the form \cite{0953-8984-28-23-235401}
\begin{equation}
\begin{split}
&p_\alpha(\mathbf{r}) \propto
 \exp\left[-\frac{\varphi(\mathbf{r}+\mathbf{r}_\alpha)}{T}-b_\alpha (\mathbf{e}_{\|\alpha}\cdot\mathbf{r})-
\right. \\
& \qquad\qquad\qquad\qquad\qquad \left.
-\frac{(\mathbf{e}_{\|\alpha}\cdot\mathbf{r})^2}{2 c_{\|\alpha}^2}-
\frac{(\mathbf{e}_{\perp\alpha}\cdot\mathbf{r})^2}{2 c_{\perp\alpha}^2}\right],
\label{RSQ-eq6}
\end{split}
\end{equation}
where $T$ is the temperature,
$\mathbf{e}_{\|\alpha}=\mathbf{r}_\alpha/r_\alpha$ is the unit vector in the direction of $\mathbf{r}_\alpha$, and
$\mathbf{e}_{\perp\alpha}$ is the unit vector in the orthogonal direction.
The normalization constant and the constants $b_\alpha$ and $c_{\|,\perp\alpha}^2$ are determined by the conditions
\begin{equation}
\begin{split}
&\int{d\mathbf{r}\;p_\alpha(\mathbf{r})}=1, \qquad \int{d\mathbf{r}\;\mathbf{r}p_\alpha(\mathbf{r})}=0, \\
&\int{d\mathbf{r}\;\mathbf{r}^2 p_\alpha(\mathbf{r})}=\sigma_\alpha^2,
\quad
\int{d\mathbf{r}\;(\mathbf{e}_{\|\alpha}\cdot\mathbf{r})^2 p_\alpha(\mathbf{r})}=\sigma_{\|\alpha}^2,
\label{RSQ-eq7}
\end{split}
\end{equation}
where $\sigma_\alpha^2=\sigma_{\|\alpha}^2+\sigma_{\perp\alpha}^2$, $\sigma_{\|\alpha}^2$, and $\sigma_{\perp\alpha}^2$ are the total MSD and its longitudinal and transverse components, which can be calculated analytically in the harmonic approximation using the Born--von Karman (BvK) approach to calculate phonon spectra\cite{Dove-book}.

In the harmonic approximation, the elemental displacement $\mathbf{u}_{\alpha,\xi}$ of the particle $\alpha$ from its equilibrium position due to the phonon $\xi$ can be written as \cite{Dove-book}
\begin{equation}
	\mathbf{u}_{\alpha,\xi}(t)=A_{\xi}\mathrm{Re}\left[\mathbf{e}_{\alpha,\xi} \exp\left({-i\left(\omega_{\xi}t-\mathbf{q}\mathbf{R}_{\alpha}\right)}\right)\right],
\label{RSQ-eq8}
\end{equation}
where $A_{\xi}$ and $\mathbf{e}_{\alpha,\xi}$ are the amplitude and polarization vector (associated with the particle $\alpha$), respectively, of the phonon $\xi$ with the wave vector $\mathbf{q}$ and frequency $\omega_{\xi}$, and $\mathbf{R}_\alpha$ is the radius vector of the unit cell, in which the particle $\alpha$ is located.
The amplitude $A_\xi$ of the phonon $\xi$ (in the classical limit) is\cite{Dove-book}
\begin{equation}
A_{\xi}=\sqrt{\frac{2k_{\mathrm{B}}T}{m N \omega^2_{\xi}}},
\label{RSQ-eq9}
\end{equation}
where $m$ is the particle mass and  $k_B$ is the Boltzmann constant.

The contribution of the phonon $\xi$ to the longitudinal component of MSDs between the $0$-th particle with $\mathbf{R}_0=0$ and the particle $\alpha$ is
\begin{equation}
 \begin{split}
& \sigma_{\|\alpha,\xi}^2=\frac{1}{\tau_\xi}\int_0^{\tau_\xi}dt\,\left(\mathbf{e}_{\|\alpha}\left(\mathbf{u}_{0,\xi}(t)-\mathbf{u}_{\alpha,\xi}(t)\right)\right)^2,
 \end{split}
\label{RSQ-eq10}
\end{equation}
where $\tau_\xi=2\pi/\omega_{\xi}$.
The transverse component $\sigma_{\perp\alpha,\xi}^2$ can be found by change $\|$ to $\perp$ in Eq.~\eqref{RSQ-eq10}.
According to Eq.~\eqref{RSQ-eq10}, we should project the eigenvectors $\mathbf{e}_{\alpha,\xi}$ (which are complex in the general case of lattices with more than one particle per unit cell) on the longitudinal and transverse directions for the particle $\alpha$
\begin{equation}
\mathbf{e}_{\alpha,\xi}=a_{\|\alpha} e^{i \phi_{\|\alpha}} \mathbf{e}_{\|\alpha}+a_{\perp\alpha} e^{i \phi_{\perp\alpha}} \mathbf{e}_{\perp\alpha}.
\label{RSQ-eq11}
\end{equation}
From Eq.~ {RSQ-eq11}, the amplitudes $a_{\|,\perp\alpha}$ and arguments $\phi_{\|,\perp\alpha}$ can be simply found as
\begin{equation}
a_{\|,\perp\alpha} = |(\mathbf{e}_{\alpha,\xi}\cdot\mathbf{e}_{\|,\perp\alpha})|, \quad \phi_{\|,\perp\alpha} = \arg{(\mathbf{e}_{\alpha,\xi}\cdot\mathbf{e}_{\|,\perp\alpha})}.
\label{RSQ-eq12}
\end{equation}
By substituting Eq.~\eqref{RSQ-eq8} into Eq.~\eqref{RSQ-eq10} and taking into account Eqs.~\eqref{RSQ-eq9}, \eqref{RSQ-eq11}, and \eqref{RSQ-eq12}, the longitudinal component of MSDs $\sigma_{\|\alpha}^2$ provided by all phonons in the lattice is obtain in the form
\begin{equation}
\begin{split}
 & \sigma_{\|\alpha}^2 = \frac{k_{\mathrm{B}}T}{m N_p}\sum_{\xi}\frac{1}{\omega^2_{\xi}}\Big[a_{\|0}^2+a_{\|\alpha}^2- \\
 & \qquad \qquad \qquad - 2 a_{\|0} a_{\|\alpha}\cos\left(\phi_{\|\alpha}-\phi_{\|0}+\mathbf{q}\mathbf{R}_{\alpha}\right)\Big],
\label{RSQ-eq13}
\end{split}
\end{equation}
where the summation should be performed over all phonons $\xi$.
To obtain the expression for the transverse component of MSDs $\sigma_{\perp\alpha}^2$, the index $\|$ should be replaced by $\perp$ in Eq.~\eqref{RSQ-eq13}.

The pair correlation function $g(\mathbf{r})$ is readily obtained using Eqs.~\eqref{RSQ-eq5}-\eqref{RSQ-eq7}, and \eqref{RSQ-eq13}.
The radial (isotropic) pair correlation function $g(r)$ is calculated by angular averaging,
$g(r)=\int{d\Omega\;g(\mathbf{r})}$, where $\Omega$ is the polar angle (in the considered 2D case).

\section{Results and Discussion}

\begin{figure}[b]
\centering
\includegraphics[width=80mm]{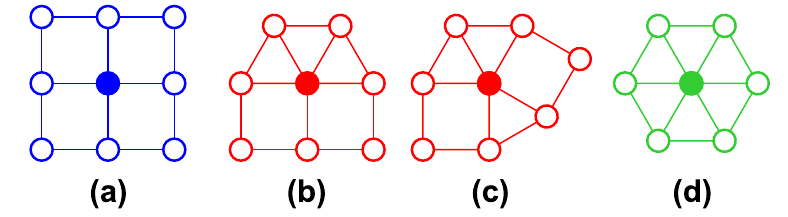}%
\caption{Possible local configurations around a (filled) particle at tiling of the plane by squares and regular triangles: (a) and (d) square and hexagonal lattices with four and six nearest neighbors, respectively;
configurations (b) and (c) provide five nearest neighbors for the filled particle.
}
\label{RSQ-Fig1}
\end{figure}

\begin{figure*}[t]
\centering
\includegraphics[width=170mm]{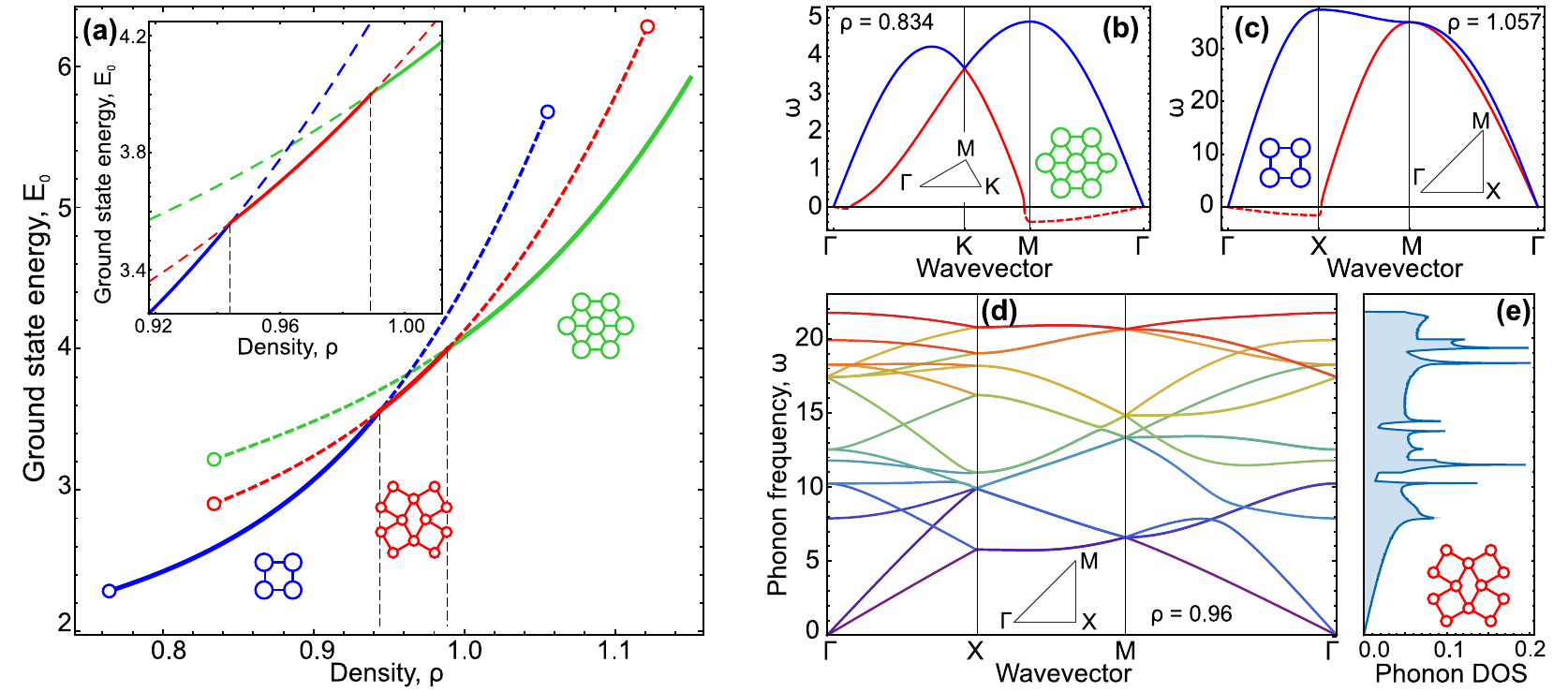}%
\caption{Thermodynamics and dynamics of the snub square lattice in the zero-temperature limit: (a) Ground state energies of the square (blue), snub square (red), and hexagonal lattices (green). The inset enlarges the region where the energy of snub square tiling is less than that in the square and hexagonal lattices. Solid and dashed lines show stable and metastable states, respectively; the circles on the lines mark the density of phonon spectrum instability. (b, c) Phonon spectra of the hexagonal and square lattices at $\rho=0.834$ and $\rho=1.057$, where the lattices become unstable (the dashed lines are the imaginary part of the frequency). (d, e) Phonon spectra and the corresponding density of states for the snub square lattice at $\rho=0.96$.}
\label{RSQ-Fig2}
\end{figure*}

\subsection{Snub square lattice}
\label{RSQ-Results-SnSq}
The interaction \eqref{RSQ-eq1} has two spatial scales associated with the core and soft shoulder diameters, which enables the stabilization of the hexagonal and square lattices at high and low densities, respectively. Geometrically, it is equivalent to the tiling of the plane by only triangles and squares. In the square lattice, the diagonal of the structure is stabilized by the soft shoulder of the potential.
However, at intermediate densities, due to the interplay between the stabilization effects at different characteristic lengths, tiling can be provided by combinations of both squares and regular triangles.
The corresponding possible combinations for local environment around particles are shown in Figs.~\ref{RSQ-Fig1}(a)-(d),
where (a) and (d) correspond to the square and hexagonal lattices with four and six neighbors, respectively, and (b) and (c) describe the possible local environment with five neighbors.
Below, it is most important that the local configurations (b) and (c) have the same interaction energy, providing by the way a \emph{doubly degenerate} state: Thermodynamically, the probabilities of these two local configurations are equal to each other if the interaction is short-range and the effect of far neighbors can be neglected.
However, contrary to the configuration \ref{RSQ-Fig1}(b), a remarkable property of the configuration \ref{RSQ-Fig1}(c) is that it can pave the plane forming the \emph{snub square lattice}\cite{CHAVEY1989147}.

\begin{figure}[!t]
\centering
\includegraphics[width=85mm]{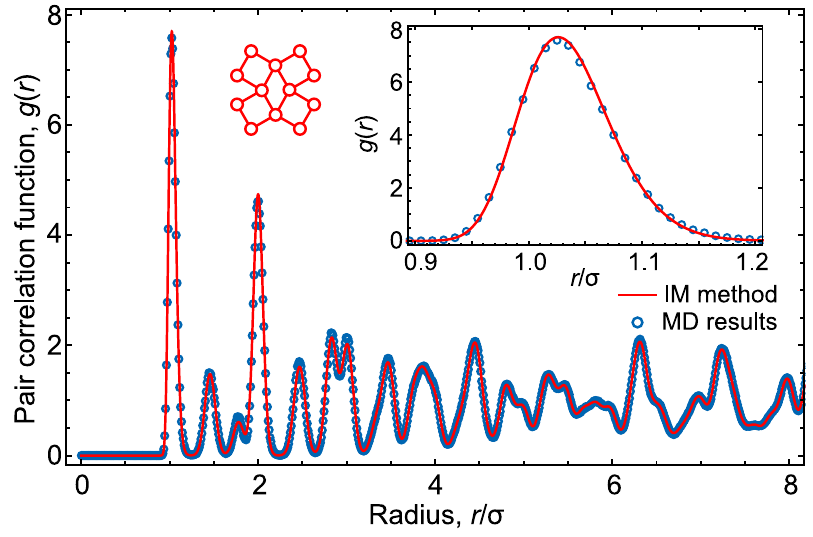}%
\caption{Pair correlation function $g(r)$ of the snub square lattice at $\rho=0.96$ and $T=0.1$. Symbols are MD results, the solid line is the completely analytical result obtained using the IM with the phonon spectra shown in Fig.~\ref{RSQ-Fig2}(d). The inset demonstrates the region of the first correlation peak.}
\label{RSQ-Fig3}
\end{figure}

We studied the snub square lattice in the zero-temperature limit; the results are shown in Figs.~\ref{RSQ-Fig2} and \ref{RSQ-Fig3}.
Figure~\ref{RSQ-Fig2}(a) presents the results for the ground state energy $E_0$ per particle calculated for the square (blue), snub square (red), and hexagonal lattices (green):
The solid lines correspond to stable states with the lowest energy and stable phonon spectra and one can see that the lowest energy in the region of $\rho=0.944$ to $\rho=0.987$ is provided by the snub square tiling of the plane.

The analysis of BvK phonon spectra allowed us to calculate the density range, where the chosen structures are dynamically stable; i.e., all squared phonon frequencies are positive.
The dashed lines in Fig.~\ref{RSQ-Fig2}(a) describe the metastable lattices (at a given density) with stable phonon spectra.
In Fig.~\ref{RSQ-Fig2}(a), we show the calculated boundaries of dynamical stability:
The circle symbols on the lines mark the densities, at which the phonon spectra become unstable even in the zero-temperature limit.
The examples of calculated phonon spectra in such points are shown in Figs.~\ref{RSQ-Fig2}(b) and (c) for the triangle and square lattices, respectively.
The dashed lines in these figures present the imaginary part of the frequency on the unstable phonon branches.
The examples of phonon spectra and corresponding density of states for the snub square lattice at the density of its static and dynamic stability are presented in Figs.~\ref{RSQ-Fig2}(d) and (e), respectively.
Since the unit cell in the snub square lattice consists of 8 particles, the spectra have 2 acoustic and 14 optical branches.

Molecular dynamics simulations show that the snub square lattice indeed does remain stable at low temperatures, $T<0.1$.
Since the spectra can be calculated using the BvK method and effects of anharmonicity are weak at low temperatures, we applied the generalized IM (see Sec.~\ref{RSQ-Methods-IM}) to compare the pair correlation functions calculated analytically and obtained using our MD simulations (details see in Sec.~\ref{RSQ-Methods-MD}).
Note that the IM method was used previously only for calculations in simple lattices (with one particle per unit cell) and has never been applied for such complex lattices as the snub square one.
Figure~\ref{RSQ-Fig3} demonstrates comparison between MD results and the completely analytical IM at the density $\rho=0.96$ and the temperature $T=0.1$.
The presented pair correlation function $g(r)$ was found using the BvK phonon spectra shown in Fig.~\ref{RSQ-Fig2}(d) and the IM generalized in Sec.~\ref{RSQ-Methods-IM}.
One can see a remarkable accuracy of the calculated $g(r)$ function in the low-temperature limit using the completely analytical IM,
whereas the resulting relative error of the excess energy $U=n/2\int{d\mathbf{r}\;g(\mathbf{r})\varphi(r)}$ is about $10^{-3}$.

\subsection{Quasicrystalline phase}
\label{RSQ-Results-HD12}

\begin{figure*}[t]
\centering
 \includegraphics[width=150mm]{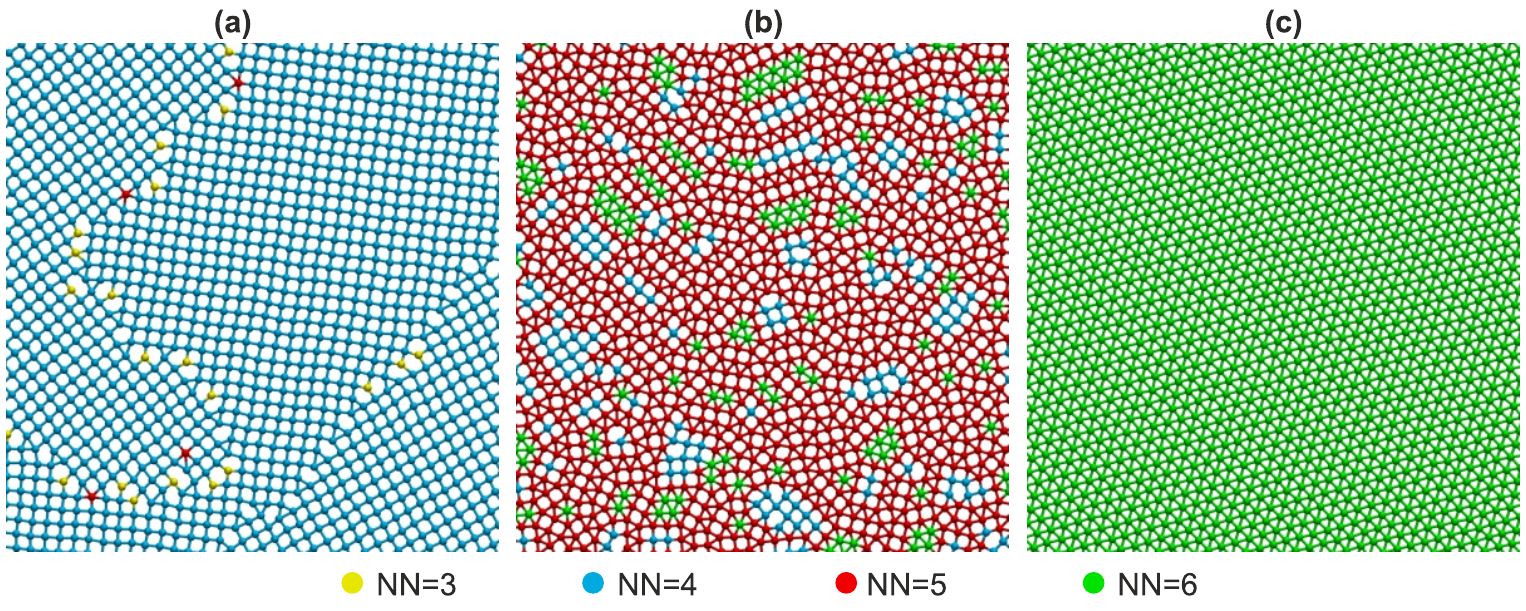}
\caption{Snapshots of the system at the temperature $T=0.12$ and densities $0.86$ (a), $0.94$ (b), and $1.02$ (c).}
\label{RSQ-Fig4}
\end{figure*}

\begin{figure*}[t]
\centering
 \includegraphics[width=160mm]{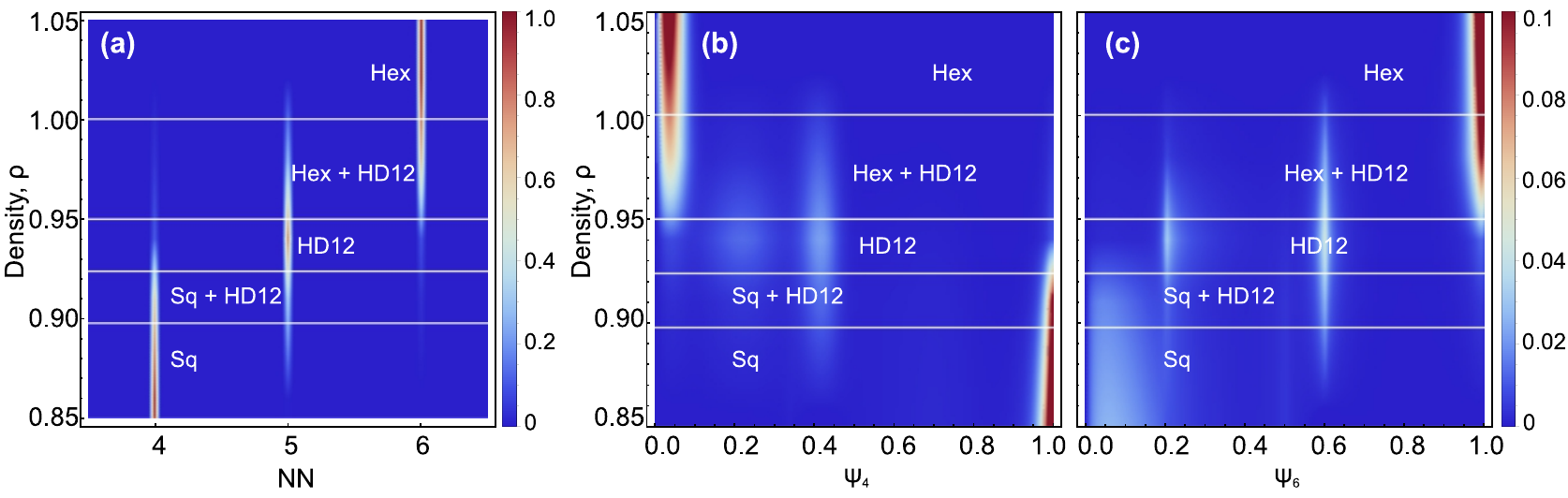}
\caption{Density dependences of the number of nearest neighbors and order parameters at low temperatures:
(a) Probability density distribution of the number of nearest neighbors $\mathrm{NN}$ (a) and the order parameters $\psi_4$ (b) and $\psi_6$ (c) at $T=0.12$ and different densities.}
\label{RSQ-Fig5}
\end{figure*}

Although the snub square lattice is stable at low temperatures,
our MD simulations show that the crystal is irreversibly broken upon heating, as is illustrated in Fig.~\ref{RSQ-Fig4}.
Snapshots of the system at densities corresponding to the stability of the square and hexagonal lattices are presented in Figs.~\ref{RSQ-Fig4}(a) and (c), respectively.
However, at the density corresponding to the (assumed) stability of the snub square lattice, we observed the generation of the lattice where a large part of the particles indeed have five nearest neighbors,
but the local environment is realized by structures shown both in Figs.~\ref{RSQ-Fig1}(b) and (c).
This situation can be explained more thoroughly using Fig.~\ref{RSQ-Fig5}, where we show the probability distributions for the number of the nearest neighbors $\mathrm{NN}$ (a), as well as the local order parameters $\psi_4$ (b) and $\psi_6$ (c) calculated at the $T=0.12$ isotherm for different densities of the system (technical details see in Sec.~\ref{RSQ-Methods-Analysis}).
One can see that most particles at a low density (e.g., $\rho=0.86$) have $\mathrm{NN}=4$, $\psi_4\approx 1$, and $\psi_6\approx0$, which clearly justifies that we observe a square lattice.
At a high density (e.g., $\rho=1.02$), we obtain $\mathrm{NN}=6$, $\psi_4\approx0$, and $\psi_6\approx1$, which corresponds to a hexagonal crystal.
However, in the intermediate range, we observe a complex behavior of the order parameters $\psi_6$ and $\psi_4$, while most particles acquire a local configuration with $\mathrm{NN}=5$.
The results shown in Figs.~\ref{RSQ-Fig4} and \ref{RSQ-Fig5} demonstrate that the transition from a square to a hexagonal lattice with an increase in the density occurs in a non-trivial manner accompanied by the generation of the local five-bond environment around particles.

\begin{figure}[h]
\centering
 \includegraphics[width=80mm]{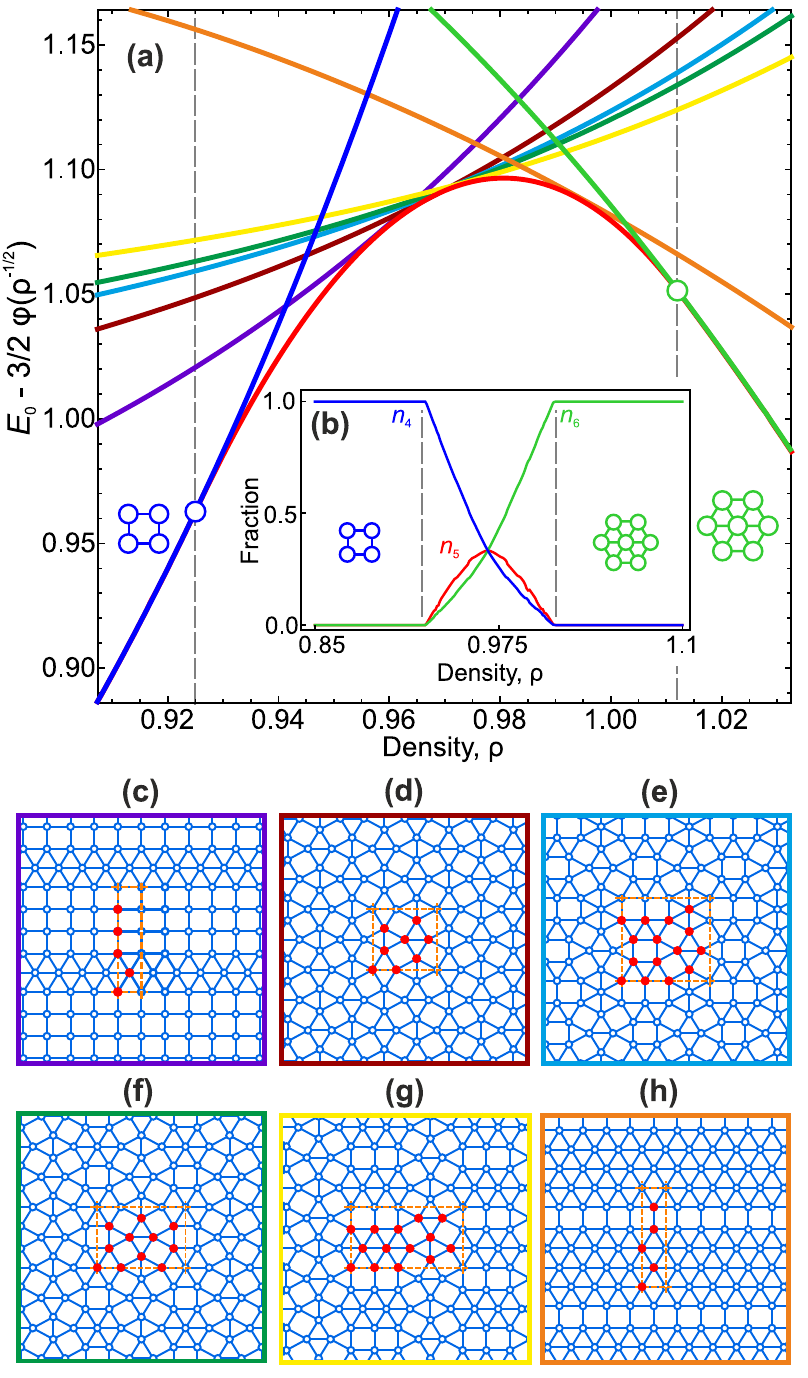}
\caption{Cascade of phase transitions at zero temperature.
    (a) Ground state energy $E_0(\rho)$ given by Eq.~\eqref{RSQ-eq17} and energies of other structures of the same density.
    (b) Fractions $n_{4,5,6}$ obtained by minimizing the total interaction energy.
    (c)-(h) Sketches of 2D complex crystals composed of squares and regular triangles, the frames are colored in accordance with curves in (a).
    The particles in unit cells are colored in red, and the cell boundaries are marked by orange dashed frames.}
\label{RSQ-Fig6}
\end{figure}

In the zero-temperature limit, the physical mechanism responsible for the observed lattice break with the generation of different local configurations can be explained qualitatively using the following approach.
Due to the short-range character of the potential \eqref{RSQ-eq1},
the total ground state energy can be calculated (in the considered density range $0.9\lesssim\rho\lesssim1.0$) as the sum of independent contributions from possible local configurations.
In terms of the distance $a$ between the neighboring particles,
the energies $E$ and areas $S$ per particle for the configurations shown Figs.~\ref{RSQ-Fig1}(a)-(d) are given by the formulas
\begin{equation}
\begin{split}
 &E_4(a)=2 \varphi(a)+2 \varphi\left(\sqrt{2}a\right), \;\; S_4 = a^2,\\
 &E_5(a)=5 \varphi(a)+2 \varphi\left(\sqrt{2}a\right), \;\; S_5 = a^2 (2+\sqrt{3})/4,\\
 &E_6(a)=3 \varphi(a), \qquad S_6 = a^2 \sqrt{3}/2,
 \end{split}
\label{RSQ-eq14}
\end{equation}
where the subscripts 4, 5, and 6 correspond to the particles with four-, five-, and six-``bond'' environment in Fig.~\ref{RSQ-Fig1}, respectively.
The total area of the system is
$$
S=N_4 S_4 + N_5 S_5 + N_6 S_6,
$$
where $N_{j}$ denotes the number of particles with the $j$-bond environment.
Hence, taking into account that $N/S = \rho$, we obtain
\begin{equation}
\begin{split}
 & a(\rho, n_4,n_5,n_6)=\rho^{-1/2}\left(n_4+n_5\frac{\left(2+\sqrt{3}\right)}{4} + n_6\frac{\sqrt{3}}{2}\right)^{-1/2},
 \end{split}
\label{RSQ-eq15}
\end{equation}
where $n_j = N_j/N$ is the concentration of the particles with $j$ bonds.
Using Eqs.~\eqref{RSQ-eq14} and \eqref{RSQ-eq15}, the energy of the system at a given density $\rho$ can be written as
\begin{equation}
	E(\rho,n_4,n_5,n_6)=\sum_{j=4}^6{n_j E_j\left(a(\rho,n_4,n_5,n_6)\right)}.
\label{RSQ-eq16}
\end{equation}

\begin{figure}[!t]
\centering
 \includegraphics[width=75mm]{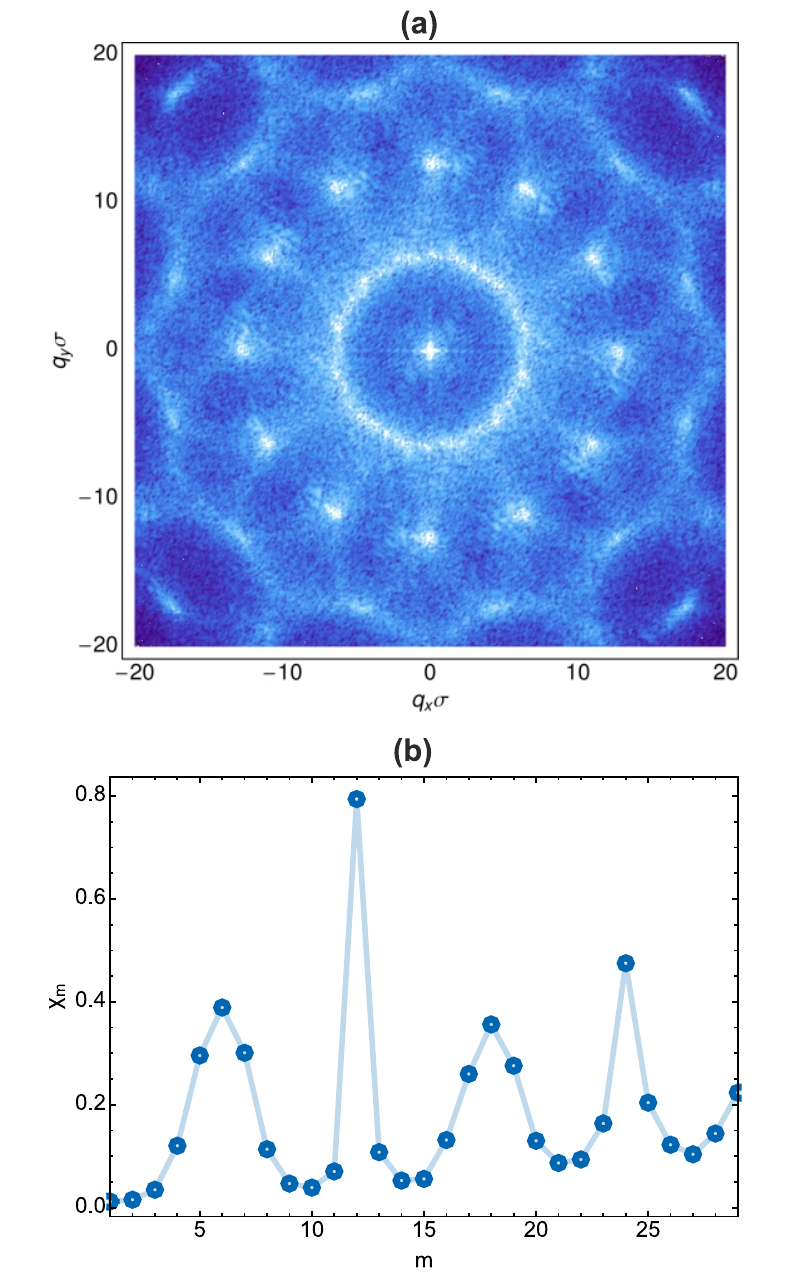}
\caption{Structure analysis of the HD12 phase: (a) the form factor $S(\mathbf{q})$ and (b) averaged order parameter $\chi_m$ calculated for the structure obtained by MD simulations at $T=0.12$ and $\rho=0.94$.
Both panels clearly illustrate the 12-fold symmetry of the observed quasicrystal.}
\label{RSQ-Fig7}
\end{figure}

The true concentrations $n_j$ can be found as non-negative values with the sum normalized to unity, which minimize the total interaction energy \eqref{RSQ-eq16}.
Hence, the ground state energy is
\begin{equation}
\begin{split}
E_0(\rho)=&\min\{E(\rho,n_4,n_5,n_6)\}, \\
 & n_4,n_5,n_6 \geq 0, \quad n_4+n_5+n_6=1.
\end{split}
\label{RSQ-eq17}
\end{equation}

Figure~\ref{RSQ-Fig6} and presents (a) the ground state energy $E_0(\rho)$ minus the value $1.5\varphi(\rho^{-1/2})$ and (b) the concentrations $n_{4,5,6}(\rho)$ of particles with different local configurations given by Eq.~\eqref{RSQ-eq17}.
The value $1.5\varphi(\rho^{-1/2})$ is subtracted from the ground state energies per particle in Fig.~\ref{RSQ-Fig6}(a) in order to distinguish more clearly the results for different structures. Apart from the energy given by Eq.~\eqref{RSQ-eq17}, dependences for several fixed crystalline structures are presented, including the square, hexagonal, snub square, and more complex lattices schematically shown in Figs.~\ref{RSQ-Fig6}(c)-(h).

One can see in Fig.~\ref{RSQ-Fig6} that the energy given by Eq.~\eqref{RSQ-eq17} for random tiling indeed corresponds to a stable state of the system, whereas the considered crystal lattices turn up to be metastable. Therefore, the transition from square to hexagonal states indeed occurs through an enormous number of random tiling lattices with smoothly-changing concentrations of particles in different local configurations.

Of coarse, the just discussed approach is approximate since it disregards a weak deformation of the squares (which become rhombi) and triangles (which loss their regularity), which is actually observed in MD simulations even at zero temperature.
Nevertheless, the explained theoretical model makes it possible to understand the physical reason and a mechanism of generating random tiling. Since the generation of random tiling is a relaxation process minimizing the corresponding thermodynamic potential of the system, it occurs asymptotically slowly at low temperatures, but becomes activated with heating of the lattice, as is justified by our MD simulations.

In accordance with Refs.\cite{Fischer01022011, Dotera2014, C6SM01454B, doi:10.1038/nmat4963}, tiling provided by squares and regular triangles can lead to the formation of a high-density quasicrystal with 12-fold symmetry (HD12 phase). For this reason, we calculated the form factor and $\chi_m$ parameter for the structures obtained in our MD simulations.
Figure~\ref{RSQ-Fig7} shows the results completely justifying the assumption about the HD12 phase. Indeed, the form factor $S(\mathbf{k})$ shown in Fig.~\ref{RSQ-Fig7}(a) is intrinsic to the HD12 quasicrystal \cite{Dotera2014} and the $\chi_m$ parameter has peaks at $m=12$ and $24$, which also indicates the quasicrystalline HD12 phase. Less expressive peaks are observed at $m=6$ and $18$ since some particles have a hexagonal local environment.

\begin{figure}[!t]
\centering
\includegraphics[width=75mm]{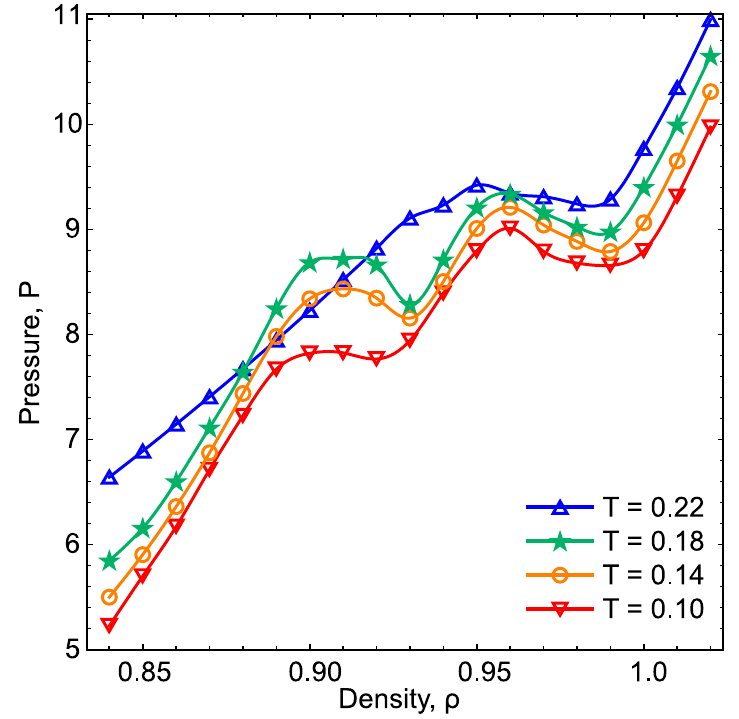}%
\caption{Equation of state of the RSP system obtained starting from a high-temperature disordered state.}
\label{RSQ-Fig8}
\end{figure}

To study the region of existence of the HD12 phase, we calculated equations of state at different isotherms, which were obtained starting from a disordered state. Since 2D systems are much less susceptible  to a glass transition than 3D ones \cite{10.1038/ncomms8392}, one can expect that the system quenching from a high-temperature configuration will rapidly fall into a stable structure under given conditions.
Figure~\ref{RSQ-Fig8} shows the equations of state of the system for several temperatures. The structure at the density $\rho=0.84$ and low temperatures is a square crystal, whereas at the highest density ($\rho=1.02$), it forms a hexagonal lattice.
One can see in Fig.~\ref{RSQ-Fig8} two sets of Mayer--Wood loops at densities about $\rho \simeq 0.9$ and $1.0$, which indicate the two first-order phase transitions from the square lattice to the HD12 quasicrystal and then from the HD12 quasicrystal to the hexagonal crystal. Using the equations of state obtained by MD simulations, one can find the transition points by the Maxwell construction.

\begin{figure}[!t]
\centering
\includegraphics[width=75mm]{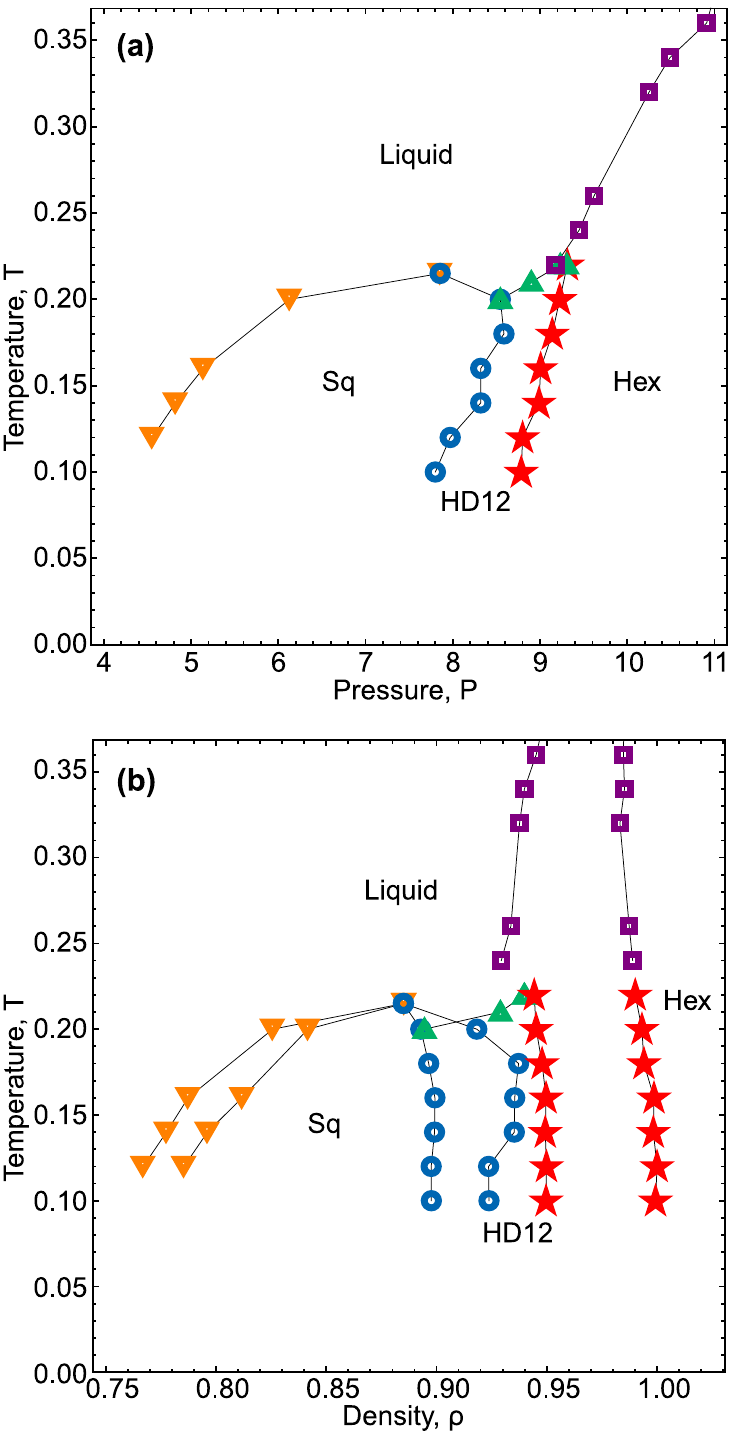}%
\caption{Phase diagram of the system in the (a) $\rho - T$ and (b) $P-T$ plane, including the areas of liquid, square (sq), hexagonal (hex) crystals, as well as the quasicrystalline phase (HD12).}
\label{RSQ-Fig9}
\end{figure}

Figure \ref{RSQ-Fig9} shows the phase diagram of the system in the region of the found HD12 quasicrystalline phase in the (a) pressure--temperature and (b) density-temperature coordinates. One can see that the HD12 quasicrystalline phase lies between the square and hexagonal lattices and exists up to the melting line. An interesting problem could be associated with the melting of HD12 quasicrystalline structures, its possible scenarios, and phase-transition cascades. However, such study is beyond the scope of this paper and should be considered in future.

\section*{Conclusion}

We have studied a 2D system of core-softened particles in an intermediate density range between the ranges corresponding to the square and hexagonal lattices. Because of two spatial scales, the considered interaction produces intermediate configurations with five nearest neighbors, and the snub square lattice appears to be more stable than the square and hexagonal lattices.
We have generalized the recently proposed IM for complex lattices and illustrated its high efficiency by employing the snub square crystal as a representative example.
However, since the five-bond environment is doubly degenerate,
this plays an important role at high temperatures leading to snub square break and quasicrystal formation, as has been observed directly in MD simulations.
Using the proposed simple qualitative model, we have illustrated the complexity of the transition between square and hexagonal lattices in the considered system even in the zero-temperature limit.
We have found that the physical mechanism responsible for this transition is associated with the generation (and subsequent annihilation) of intermediate five-bond configurations with an increase in the density.
By the way, the transition from square to hexagonal lattices is accompanied by continuously changing concentrations of realized local configurations with different numbers of the nearest neighbors.
The analysis of generated structures has shown that random tiling by squares and triangles results in a high-density quasicrystalline phase with 12-fold symmetry.
Equations of state for the system at different isotherms demonstrate Mayer--Wood loops, which indicate first-order phase transitions.
Following Maxwell reconstruction, we calculated the boundaries of the HD12 phase between the square and hexagonal lattices.
According to the phase diagram obtained, the HD12 state exists at intermediate densities and pressures and can be prepared from the square lattice by isotropic compression.

The found reversible transitions between different lattices open interesting prospects for applications of the results obtained in this work. For instance, additionally to the known method of chemical self-assembly of colloidal quasicrystals \cite{ADFM:ADFM201002091},
colloidal (polymer) particles with a soft external core can be assembled in external rotating electric fields, as has recently been done in Refs.\cite{s41598-017-14001-y, doi:10.1021/acs.jpcc.7b09317}.
External electric fields, rotating at frequency much higher than the inverse Brownian time of colloidal particles, induce tunable attraction between particles, which can be employed to increase the pressure in two-dimensional clusters.
By the way, core-softened colloidal particles can be compressed and reversible transitions between square, HD12 quasicrystal, and hexagonal phases can be induced.
Such colloidal quasicrystals with a tunable structure (and properties) can be used in a broad range of issues, from sensing to optical frequency transformations, plasmonic devices, metamaterials, and photonic crystals.
Moreover, colloidal many-body systems can be visualized for corresponding particle-resolved studies.
Such applied and fundamental studies should enable better understanding the kinetics and dynamics of quasicrystal formation at the level of individual particles and should be of interest for future experimental studies.
Since interactions with two spatial scales are widespread, from molecular systems to colloidal suspensions, we believe that the results of this study will be useful for a broad range of problems in physical chemistry, materials science, and soft matter.

\section*{Acknowledgments}
We are grateful to Acad. V.V. Brazhkin, E.E. Tareyeva, and Dr. K.A. Komarov for stimulating discussions. We used computing resources of the federal collective usage center Complex for Simulation and Data Processing for Mega-Science Facilities at the National Research Center "Kurchatov Institute", http://ckp.nrcki.ru and supercomputers at the Joint Supercomputer Center of the Russian Academy of Sciences (JSCC RAS).
The molecular dynamics simulations were supported by the Russian Foundation for Basic Research Grant No. 17-02-00320. The analysis was supported by the Russian Science Foundation Grant No. 17-19-01691.

\bibliography{Ref-RSQ}
\bibliographystyle{rsc} 

\end{document}